\documentclass[preprint,jgrga]{agu2001}

\ifgalley
\let\eject\relax
\fi

\usepackage{graphicx}
\setkeys{Gin}{draft=false}
\usepackage{latexsym}

\titlerunninghead{CMI Growth Rates }
\authorrunninghead{Mutel et al.}

\authoraddr{R. L. Mutel, Dept. Physics and Astronomy, University of Iowa,
Iowa City IA 52242 (robert-mutel@uiowa.edu)}
\authoraddr{W. M. Peterson, Dept. Physics and Astronomy, University of Iowa,
Iowa City IA 52242 (william-peterson@uiowa.edu)}
\authoraddr{T. R. Jaeger, Dept. Physics and Astronomy, University of Iowa,
Iowa City IA 52242 (theodore-jaeger@uiowa.edu)}
\authoraddr{J. D. Scudder, Dept. Physics and Astronomy, University of Iowa,
Iowa City IA 52242 (jack-scudder@uiowa.edu)}

\begin{document}
\def\etal{et al.\ }
\def\jgr{{\it J. Geophys. Res.}}
\def\grl{{\it Geophys. Res. Lett.}}
\def\apj{{\it Astrophys. J.}}
\def\npg{{\it Nonlin. Proc. Geophys.}}

\title{Dependence of CMI Growth Rates on Electron Velocity Distributions
and Perturbation by Solitary Waves}
\authors{
R. L. Mutel, \altaffilmark{1}
W. M. Peterson, \altaffilmark{1}
T. R. Jaeger,  \altaffilmark{1}
J. D. Scudder,  \altaffilmark{1}
}

\altaffiltext{1}
{Dept. Physics and Astronomy, University of Iowa, Iowa City IA 52242, USA.}

\begin{abstract}
We calculate growth rates and corresponding gains for RX and LO mode radiation associated with
the  cyclotron maser instability for parameterized horseshoe electron
velocity distributions. The velocity distribution function was modeled to closely fit the electron distribution functions observed in the auroral cavity. We systematically varied the model parameters as well as the propagation direction to study the dependence of growth rates on model parameters. The growth rate depends strongly on loss cone opening angle, which must be less than 
$90^{o}$  for significant CMI growth. 
The growth rate is sharply peaked for perpendicular radiation ($k_{\parallel}\ =\ 0$), with a full-width at half-maximum $1.7^{o}$, in good agreement with observed k-vector orientations and numerical simulations. The fractional bandwidth varied between 10$^{-4}$ and 10$^{-2}$, depending most strongly on propagation direction. This range encompasses nearly all observed fractional AKR burst bandwidths. We find excellent agreement between the computed RX mode emergent intensities and observed AKR intensities assuming convective growth length  $L_c\approx$20-40 km and group speed 0.15$c$.  The only computed LO mode growth rates compatible observed LO mode radiation levels occurred for number densities more than 100 times the average energetic electron densities measured in auroral cavities. This implies that LO mode radiation is not produced directly by the CMI mechanism but more likely results from mode conversion of RX mode radiation. We find that perturbation of the model velocity distribution by large ion solitary waves (ion holes) can enhance the growth rate by a factor of 2-4. This will result in a gain enhancement more than 40 dB depending on the convective growth length within the structure.   Similar enhancements may be caused by strong EMIC waves.
\end{abstract}

\begin{article}
\section{Introduction}
The electron cyclotron maser instability (CMI) is widely recognized as the
generator of auroral kilometric radiation (AKR) from the Earth's
magnetosphere (see \citeauthor{t06} (\citeyear{t06}) for a recent review). It is also been invoked to explain radiation
from many other astrophysical sources, including Jovian magnetospheres
\citep{z98}, microwave bursts from certain solar bursts \citep{md82,dw87},
flare stars \citep{bb87}, white-dwarf binaries
\citep{w04}, and even extragalactic radio jets \citep{b05}.  
The free energy of the maser derives from mildly energetic downward-directed electron beams which are accelerated by parallel electric fields and pitch-angle scattered by magnetic field gradients at lower altitude. The resulting distribution function, sometimes referred to as a 'shell' or 'horseshoe' distribution, has a large crescent-shaped shoulder of positive velocity gradient ($\partial f/ \partial v_{\bot} > 0$), resulting in strong non-linear growth of waves near the electron cyclotron frequency. The dominant free-space emission is the right-hand circularly polarized extraordinary (R-X) mode, whose spectral power has been measured as large as $10^{-4}$ V$^{2}$ m$^{-2}$ Hz$^{-1}$ \citep{e98} for bursts with typical bandwidth of one KHz. This is equivalent to an energy flux $2\cdot 10^{-4}$ W m$^{-2}$ , which is $\sim$1\% of the kinetic power of the electron beam, a very high efficiency since many of the beam electrons precipitate to the ionosphere and do not contribute to the CMI instability. 

Early papers which identified AKR emission with a CMI mechanism often assumed a loss-cone electron velocity distribution (e.g. \citet{w79, lw80,w82,og82, owg84}).  A key insight was that the relatively small relativistic mass correction applied to the cyclotron resonance condition resulted in a resonance ellipse rather that a straight line in velocity space \citep{w79}. The advantage of the ellipse is that it could intercept a significant fraction of the  'inverted' ($\partial f/\partial v_{\bot}>0$) regions of the loss-cone.   However, \citet{og84}, using S3-3 electron distributions and ray-tracing models of effective path length, showed that the computed path-integrated growths were too low to account for observed AKR intensities. Another problem with loss-cone models is that in order for the resonant ellipse to intercept the 'shoulders' of the loss cone (where $\partial f / \partial v_{\bot} > 0$), the ellipse must be significantly offset from the origin, which in turn requires the k-vector to have a significant upward directed parallel component. This is in conflict with in situ AKR observations in the auroral cavity, which indicate that the k-vector is perpendicular to the magnetic field direction within a few degrees \citep{e98}. The loss-cone model became even more problematic when \citet{l90} reported that Viking observations of electron distributions in the auroral cavity having loss-cones with large perpendicular velocity gradients were often not associated with any AKR emission. 

A second key theoretical insight was that relativistic corrections also play a crucial role in the dispersion relation, even for mildly relativistic electrons \citep{p84a, p84b,s86}.  In the low-density environment of the auroral cavity, the relativistic correction causes the RX-mode cutoff frequency to be shifted below the cyclotron frequency.  For nearly normal  propagation, this produces a new relativistic mode which is highly unstable \citep{p84b}. It is this mode which dominates CMI growth.

In the mid-1980's, detailed simulations of the CMI mechanism lent strong support to the idea that the growth rate was driven by an electron distribution function with a shell or horseshoe geometry rather than a loss-cone \citep{ps85, wp86}. This was confirmed in the early 1990's when in situ measurements  indicated that the distribution function in auroral cavities does indeed often resemble a horseshoe shape \citep{r93, d98,e98,s01}. The horseshoe shape is a consequence of Earthward-directed beams of 1-10 keV electrons which are accelerated by parallel electric fields and pitch-angle scattered by the converging magnetic field.  The origin of the parallel electric field is controversial, but may be a result of mid-cavity double layers \citep{e04}. The horseshoe distribution results in much more robust CMI growth since the resonant ellipse can be centered on the origin ($k_{\|} \sim 0$) and hence can intersect a much larger volume of velocity space with a positive velocity gradient. This realization has resulted in a number of papers calculating CMI growth rates using specific horseshoe distributions \citep{bc00,p02,rlm06}.

Both left-hand ordinary circularly polarized (L-O mode) and R-X mode emission are detected by remote spacecraft. Internal Z mode is also detected in the source region \citep{g83}, but may result from conversion of RX-mode radiation propagating normal to the auroral cavity \citep{p02}.  Although the R-X mode almost always dominates observed AKR emission, there is often significant power in the LO mode, usually in the range 1\% - 10\% of the RX power  \citep{m84,b88,h03}.  When L-O and R-X mode are seen together, their observed fluxes are often highly correlated, with L-O mode flux densities about 2\% of RX-mode \citep{m84}. This suggests that LO-mode could result from mode conversion of R-X mode radiation, perhaps by reflection at sharp density boundaries \citep{c82,hm86}. More recently, \citet{h03} reported that L-O mode AKR is detected much more often at low magnetic latitudes ($\lambda_{m}\sim 60^{o}$) in the afternoon auroral sector compared with the evening sector. They attributed this to a propagation effect: R-X mode originates in the evening sector where AKR bursts are most often located \citep[e.g.,][] {h88,rlm03,rlm04}. However, some fraction of RX mode radiation is mode converted to LO mode (e.g. by reflection from density boundaries), and propagates along largely unrefracted ray paths over the polar cap, allowing some spacecraft locations to intercept  only L-O mode ray paths. This idea was first suggested by \citet{h84} and was also seen in DE-1 spacecraft observations of R-X and L-O mode AKR \citep{m84}.

The question of whether L-O mode AKR is primarily generated directly by the CMI mechanism, or whether it is mode-converted from RX-mode, remains an unresolved issue. Since the CMI growth rate is proportional to number density, in regions of relatively high density, L-O mode is favored since the R-X mode is strongly suppressed by even modest $\omega_{pe}/\Omega_{ce}$ ratios. In the limiting case of no cold electrons, the relativistic RX-mode cutoff frequency is
\begin{equation}
\omega_{rx} = \frac{ \Omega_{ce}}{2\gamma} +
{\left[ {\left( \frac{\Omega_{ce}}{2\gamma}\right)}^{2}+{\omega_{pe}}^{2}\right]}^{1/2}
\end{equation}
 where $\gamma$ is the Lorentz factor of energetic electrons. This cutoff exceeds 
the maximum allowable frequency for CMI growth ($\Omega_{ce}$) at very modest densities. For example, for 5 keV electrons ($\gamma = 1.01$), CMI generation of RX-mode is suppressed for $\omega_{pe}/\Omega_{ce} > 0.1$, corresponding to a limiting density $n_{e} = 3 $\ cm$^{-3}$ at frequencies below 225 kHz. Since the cutoff frequency for LO-mode is $\omega_{pe} \ll \Omega_{ce}$ in the lower magnetosphere, it can continue to grow linearly with increasing density if the CMI resonance ellipse (whose topology depends in part on density) intercepts a suitable region of the velocity distribution.

AKR exhibits a bewildering variety of fine structure in dynamic (frequency-time) spectra. \citet{ga81} first suggested that the drifting narrowband features were the result of a propagating disturbance which triggers narrowband AKR emission at the local cyclotron frequency. Several recent papers ascribe AKR fine structure to solitary structures such as electron holes \citep{ptbj03}, ion holes \citep{rlm06}, tripolar structures \citep{pt05}, or to electromagnetic ion-cyclotron waves \citep{men06,x07}. These waves propagate along (or obliquely to) field lines, and are thought to modify the electron distribution so as to enhance the CMI growth rate. \citet{rlm06} calculated the effect of ion holes on a horseshoe electron and found that the resulting R-X mode CMI emitted power was enhanced by $\sim$20 dB. Both the magnitude and the narrowband nature of the of the enhancement were consistent with drifting 'striated' AKR bursts reported by \citet{men96,men00}. 

In this paper we investigate three questions stimulated by the observational and theoretical investigations of AKR summarized above:

$(1)$ How does the CMI growth rate vary with horseshoe parameters and with propagation direction and ambient density? More particularly, under which conditions is the growth a maximum?   To answer this, we performed a systematic parameter search, varying the electron velocity distribution parameters, the ratio of electron plasma to cyclotron frequency, and wave-normal angle to the ambient magnetic field. 

$(2)$ What is the ratio of growth rates for R-X mode compared with L-O mode for horseshoe distributions with the same range of parameters? Are there parameters for which direct L-O mode growth rates reproduce the observed flux range of 1\% - 10\% of R-X mode power? Is there a significant difference in maximum growth rate k-vector orientation for LO-mode vs. R-X mode which may be relevant to the observations of differing RX/LO ray paths?

$(3)$ Under what conditions can solitary structures enhance (or diminish) the CMI growth rate as a result of the modification of the electron distribution function interior to the structure?

We  address these questions by calculating the growth rates and emergent spectral power for a parameterized horseshoe electron velocity distribution. The distribution function has been designed to closely fit the observed a canonical velocity distribution in the Earth's auroral cavity. By varying the parameters centered on the best-fit values to the  observed distribution, as well as the propagation angle, we investigate how growth rates vary with each parameter and propagation direction, and how the growth rates are modified by perturbations . We consider only fundamental (first harmonic) mode emission and do not consider generation of the trapped Z mode, or mode conversion processes. We also assume a steady-state process in which the input electron beam is constant and hence do consider not the evolution of the electron distribution function after interaction with the radiation field. Finally ,we assume that the fraction of cold electrons in the auroral cavity is very small, so that the energetic component dominates the total electron density.

\section{CMI growth rates: Parameter search}

\subsection{CMI growth rate equations}
Growth rates for the mildly relativistic Doppler-shifted cyclotron resonance condition can be written \citep{w79,w85}:

\begin{equation}
\omega _i  = \mu_R\frac{{\pi ^2 \omega _{pe} ^2 }}{{4n_e }}\int_{-\infty}^\infty  {dv_\parallel  } \int_0^\infty  {dv_ \bot  } v_ \bot  ^2  \delta \left( {v_\parallel  ,v_ \bot  } \right)\frac{{\partial f}}
{{\partial v_ \bot  }}
\end{equation}

for the right-hand polarized extraordinary (R-X) mode, and

\begin{equation}
\omega _i  = \mu_L\frac{{\pi ^2 \omega _{pe} ^2 }}{{4n_e c^2 }}\int_{-\infty}^\infty  {dv_\parallel  } \int_0^\infty  {dv_ \bot  } v_ \bot  ^2 v_\parallel  ^2 \delta \left( {v_\parallel  ,v_ \bot  } \right)\frac{{\partial f}}{{\partial v_ \bot  }}
\label{eqn-dftn}
\end{equation}

for the left-hand polarized ordinary (L-O) mode. In both equations $n_e$ is the number density of energetic electrons, $\omega_{pe}$ is the plasma frequency, $f$ is the normalized velocity distribution function (so that the integral of $f$ over velocity is the number density), and we have assumed $\omega_{r}\sim\Omega_{ce}$. We also assume nearly perpendicular propagation  ($k_{\|}\ll k$) and hence omit the term $\frac{k_{\|}}{k} \frac{\partial f} {\partial v_{\|}}$ since it is insignificant compared with the $\partial f /\partial v_{\bot}$  term.  The dimensionless factors $\mu_{R,L}$  correct for the effect of plasma dispersion. This was ignored in the original derivation of \citet{w79}, who assumed a refractive index $n= 1$. The values of $\mu_{R,L}$ are discussed in Appendix A.

The delta function is given by the Doppler-shifted gyro-resonance condition
\begin{equation}
\delta \left( {v_\parallel  ,v_ \bot  } \right) = \delta \left\{ {\omega _r  - \frac{{\Omega _{ce} }}{{\gamma \left( {v_ \bot  ,v_\parallel  } \right)}} - k_\parallel  v_{\|} } \right\}
\end{equation}
 
For $v \ll c$ this equation can be rewritten in the useful form
\begin{equation}
\delta \left( {v_\|  , v_{\bot} } \right) = \delta \left\{ {\frac{{\Omega _{ce} }}{{2c^2 }}\left[ {v_ \bot  ^2  + \left( {v_\parallel   - v_c } \right)^2  - v_r ^2 } \right]} \right\}
\end{equation}

The argument of the delta function is the equation of a circle whose center is displaced along the $v_{\|}$ axis by
\begin{equation}
v_c = \frac{{k_\parallel  c^2 }}{{\Omega _{ce} }}
\label{eqn-offset}
\end{equation}

and whose radius is

\begin{equation}
v_r  = \sqrt { {v_c }^2  - 2c^{2}\delta \omega } 
\label{eqn-radius}
\end{equation}

where $\delta\omega = \left(\omega_{r} -\Omega_{ce}\right)/\Omega_{ce}$.  The growth rate equations (2) and (3) can be recast by converting to polar coordinates, resulting in

\begin{equation}
\left. {\frac{{\omega _i }}{{\Omega _{ce} }}} \right|_{rx}=\mu_R\frac{{\left( {\pi v_r c} \right)^2 }}{{4n_e }}\left( {\frac{{\omega _{pe} }}{{\Omega _{ce} }}} \right)^2 \int_{0}^{\pi} {d\phi \sin^{2}\left(\phi\right) \left. {\frac{{\partial f\left( {\vec v} \right)}}{{\partial v_ \bot  }}} \right|_{\left| {\vec v} \right| = v_r } } 
\label{grow-rx}
\end{equation}

and

\begin{equation}
\left. {\frac{{\omega _i }}{{\Omega _{ce} }}} \right|_{lo}=\mu_L \frac{{\pi ^2 v_r ^4 }}{{16n_e }}\left( {\frac{{\omega _{pe} }}{{\Omega _{ce} }}} \right)^2 \int_{0}^{\pi} {d\phi \sin^{2}\left(2\phi\right) \left. {\frac{{\partial f\left( {\vec v} \right)}}{{\partial v_ \bot  }}} \right|_{\left| {\vec v} \right| = v_r } } 
\label{grow-lo}
\end{equation}

where the integration is performed around the resonant semi-circle in ($v_{\|},v_{\bot}$) velocity space. This representation is useful in showing that the dependence of the growth rate on the velocity distribution is a linear function of the perpendicular derivative weighted by a simple geometrical factor centered at 90$^{o}$ (R-X mode) or 45$^{o}$ and 135$^{o}$ (LO mode). 

\subsection{Model velocity distribution function}
\begin{figure*}[p]
  \vspace*{2mm}
  \begin{center}
    \includegraphics[width=6.5in,angle=90]{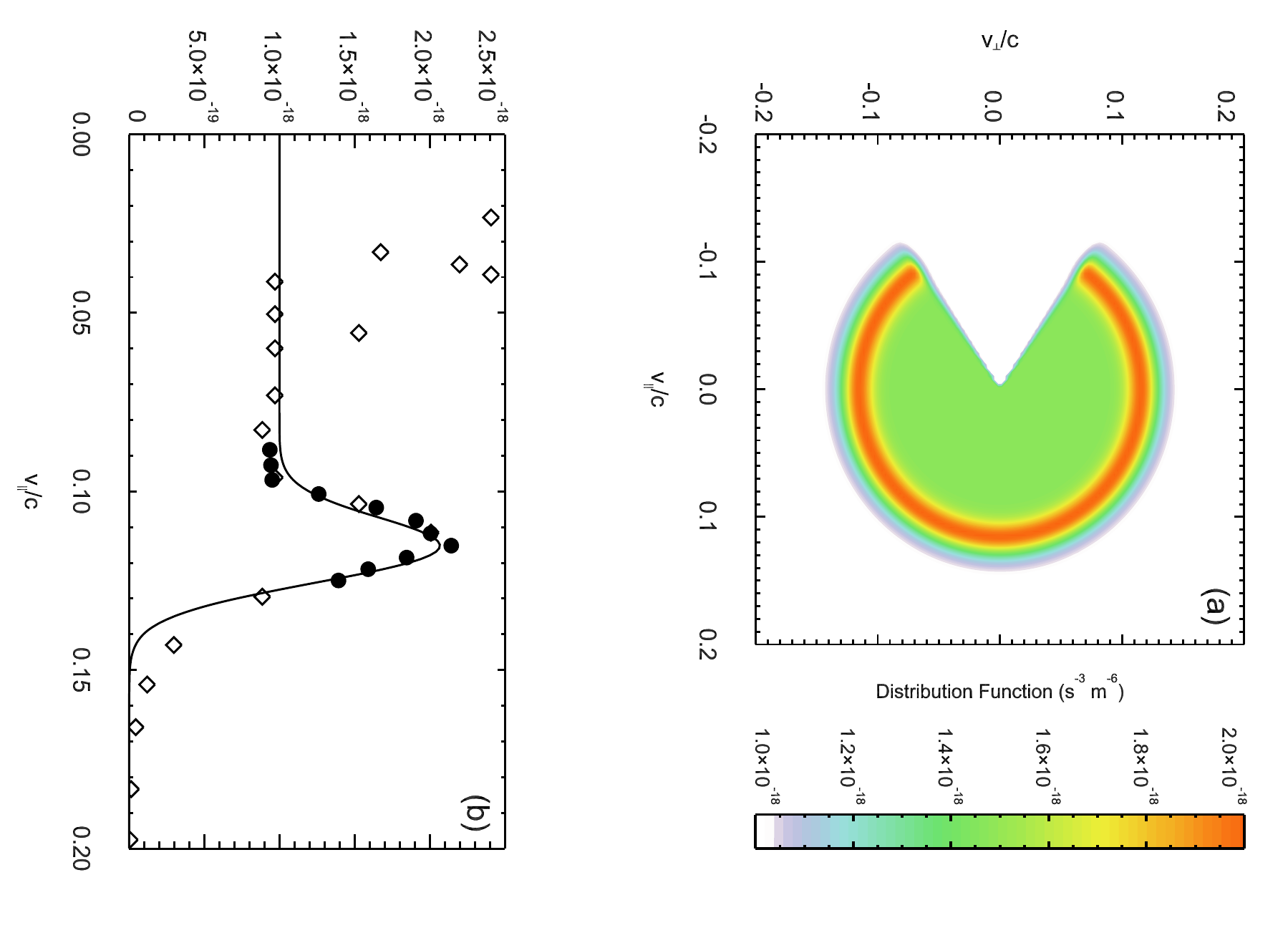}
  \end{center}
  \caption{
$(a)$ Horseshoe model electron velocity distribution function with parameters providing a best-fit to  measured distribution function described in \citet{m03} with parameters listed in Table 1 (FAST match).  $(b)$ One-dimensional cut outside the loss cone, compared with measured distribution from \citeauthor{m03}, Fig. 6 (open diamonds) and Fig. 11 (lowest trace, filled circles). Note that the distribution functions in \citeauthor{m03} have been scaled to velocity units.
  }
   \label{fig-2x1-vdf}
\end{figure*}

We modeled the velocity distribution as a partial annulus (horseshoe) with five adjustable parameters: Scale parameter $F_{0}$ (proportional to the number density) , radius ($v_r$), characteristic width ($\sigma$), loss-cone opening angle $\theta$ in the upward (anti-Earthward) direction, and horseshoe contrast parameter $\eta$.  The functional form of the assumed velocity distribution function is

\begin{eqnarray}
f \left( v_{\bot}  ,v_{\parallel} \right)  &=&
F_{0}
\left\{
\frac{1}{2}{\rm erfc}\left(
 \frac{\left| {\vec v} \right| - v_r }{\sigma } \right)
\; + 
\right. \nonumber \\ & & \! \left.\rule{0in}{3.0ex}
\eta\cdot {\rm exp}{ \left[-\left( {\frac{{\left| {\vec v} \right| - v_r }}{\sigma }} \right)^2 \right]}
\right\}\times g\
\end {eqnarray}
where the loss cone truncation function $g$ is
\begin{eqnarray*}
g = \left\{ 
\begin{array}{ll}
1 & \mbox{if $\left|\rm{atan2}\left(-v_{\bot} ,-v_{\parallel} \right)\right|>\theta$} \\
0 & \mbox{otherwise}
\end{array}
\right.
\end{eqnarray*}

where atan2 is the quadrant-dependent inverse tangent function. The velocity $|\vec{v}|$ is the quadrature sum of $v_{\bot}$ and $v_{\parallel}$, and erfc is the complementary error function. The velocity distribution function is normalized such that the integral over velocity space is equal to the electron number density
\begin{eqnarray}
n_{e} &=& \int_{-\infty}^{\infty}d^{3}\vec{v} f(\vec{v}) = 
\nonumber \\ & & \qquad \qquad
2\pi\int_{-\infty}^{\infty}dv_{\parallel}\int_{0}^{\infty}dv_{\bot}v_{\bot}
f\left(v_{\bot},v_{\parallel}\right)
\label{eqn-ne}
\end{eqnarray}
The form of the model distribution function, consisting of narrow horseshoe population superposed on a flat plateau,  was chosen to match observed velocity distributions in the auroral cavity.  Fig. \ref{fig-2x1-vdf}$(a)$ shows a pseudo-color plot of a model distribution function, while Fig. \ref{fig-2x1-vdf}$(b)$ shows a one dimensional radial profile comparing an observed distribution sampled in the auroral cavity by the FAST spacecraft \citep{m03} The open diamonds are from Fig. 6 upper panel, FAST orbit 1804, averaged over $0^{o} - 68^{o}$ pitch angle and converted to velocity units. The closed circles are from \citeauthor{m03} Fig. 11 (lowest trace), FAST orbit 11666. Although the data points are measurements made at different epochs, they were both observed while FAST was inside the auroral cavity and agree with each other and with the model distribution very well. Note that we have not fitted the low-energy electron population ($E\leq100$ eV) since most are likely photoelectrons and secondaries generated by the spacecraft \citep{d98}. 
The loss cone width angle was fitted to the electron beam depletion evident between about $140^{o} - 220^{o}$ pitch angle [\citeauthor{m03}, Fig. 6, lower panel].  

The best-fit model parameter values corresponding to the observed distribution described above are given in the column labeled 'FAST match' in Table 1. The corresponding electron number density, as given by integrating over the distribution function (equation \ref{eqn-ne}), is 0.32 cm$^{-3}$, which is in exact agreement with the measured density reported by \citeauthor{m03}

\subsection{Parameter search}
In addition to the five parameters associated with the model velocity distribution, we also varied the propagation direction (parallel wavenumber ratio $k_{\|}/k$). We varied all six parameters over the ranges listed in Table 1. For each set of  parameters, the growth rate integrals were evaluated numerically using Ridder's method for the partial derivatives  and Romberg's method for the integrations \citep{p95}. In addition, the allowed frequencies between $\Omega_{ce}$ and the RX or LO cutoff frequency were divided into 30 intervals and the maximum growth rate was recorded. The computation of a complete parameter search over the allowed range of all variables resulted in computation of growth rates for more than 240 million combinations of parameters.   In order to determine which velocity distributions and ambient conditions resulted in the largest growth rates, we sorted the output datasets separately by RX and LO growth rates. 

\subsection{Frequency dependence of growth rate}
\begin{figure*}[p]
  \vspace*{2mm}
  \begin{center}
    \includegraphics[width=8in,angle=90]{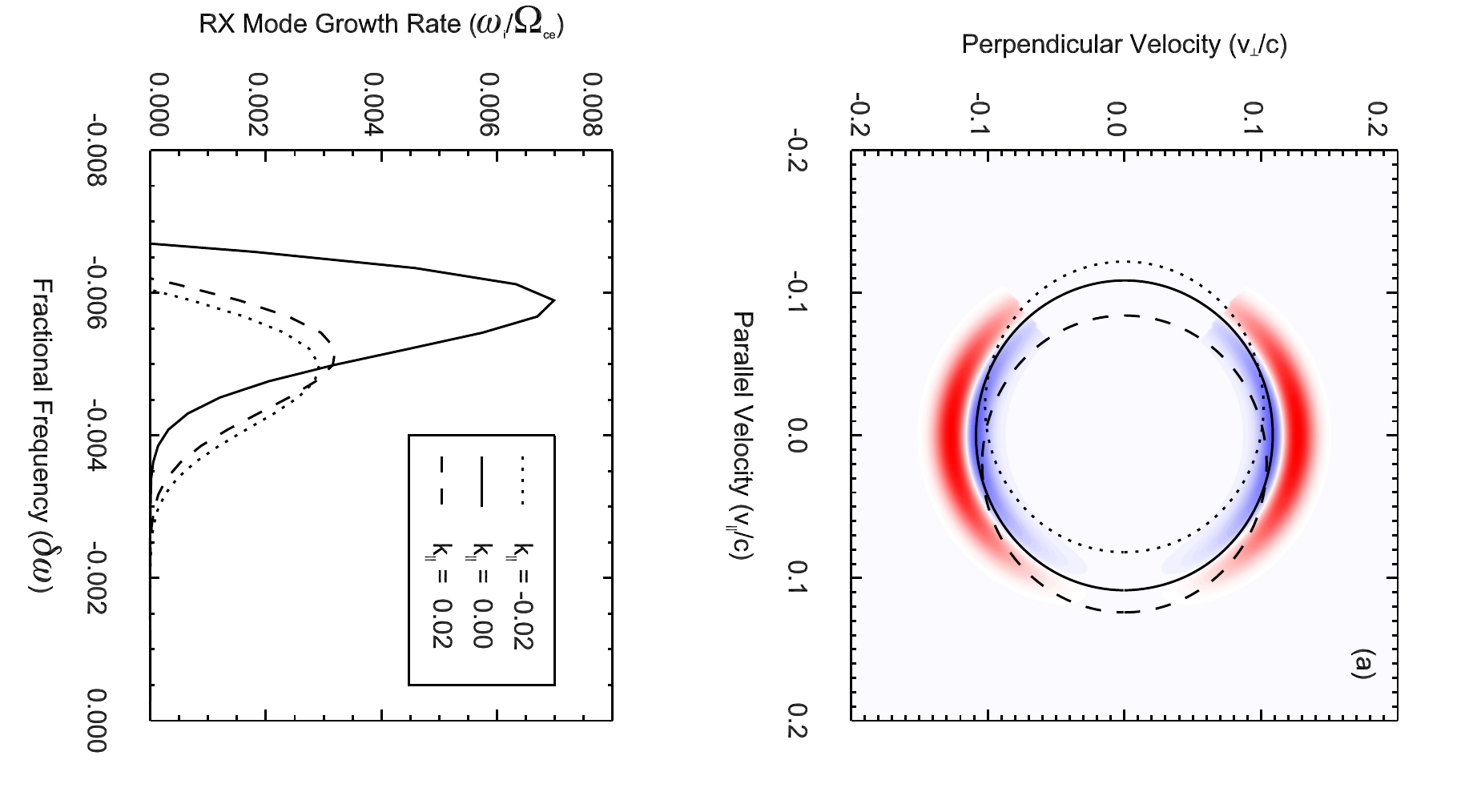}
  \end{center}
  \caption{$(a)$ Pseudo-color representation  (blue $>0$; red$< 0$) of R-X mode growth rate integrand ($sin^{2}\phi \times \partial f / \partial v_{\bot}$) for distribution function shown in Fig. 1. Resonant circles are shown for maximum growth rate at
  propagation directions $k_{\|}/k = -0.02\ (-1.1^{o}$, dashed line),  and $k_{\|}/k = 0.0\ (0^{o}$, solid line), 
   $k_{\|}/k = +0.02\ (+1.1^{o}$, dotted line). $(b)$
  R-X mode growth rate vs. fractional frequency  for propagation directions given in $(a)$. }
 \label{fig-rx-kpar}
\end{figure*}

Fig. \ref{fig-rx-kpar}$(a)$ shows a pseudo-color plot of the R-X mode growth rate integrand (equation \ref{grow-rx}) for the model distribution shown in Fig. 1, along with resonant circles corresponding to maximum growth rates for perpendicular propagation (solid circle), $k_{\|}/k$ = -0.02 ($-1.1^{o}$), and $k_{\|}/k$ = +0.02 ($1.1^{o}$). Fig \ref{fig-rx-kpar}$(b)$ shows the corresponding growth rates as a function of fractional frequency 
\begin{equation}
\delta\omega = \frac{\omega - \Omega_{ce}}{\Omega_{ce}}
\end{equation}
The narrow-band nature of CMI emission is evident, as is the strong dependence of growth rate on propagation angle (discussed further in section 2.6). The fractional bandwidth, which we can define as the full width to half-maximum, is about 0.002. This is in close agreement with the model calculations of\citep{y98} who used a similar analytic distribution function, and with the numerical simulations of \citep{p99}. 

\begin{figure*}[p]
  \vspace*{2mm}
  \begin{center}
  \includegraphics[width=5.5in,angle=90]{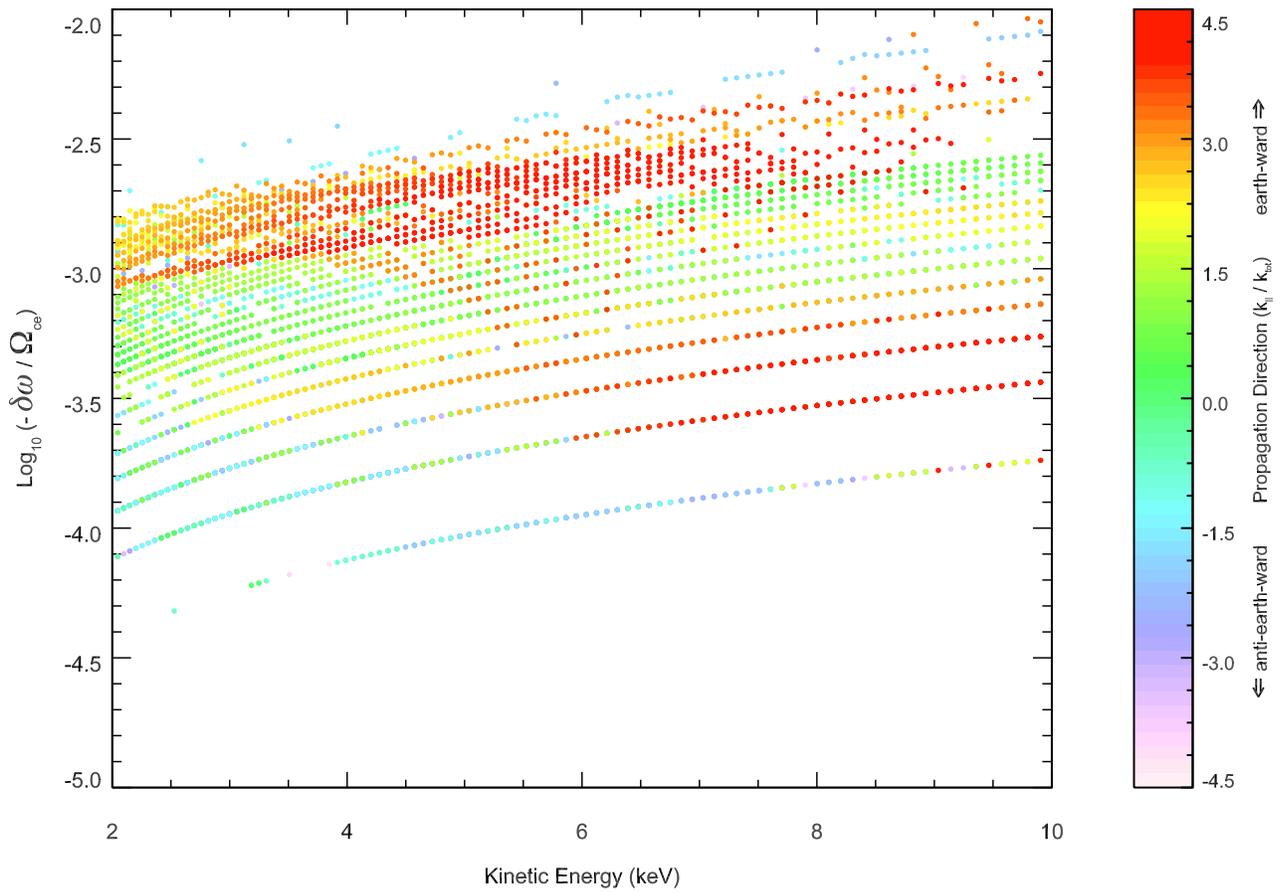}
   \end{center}
  \caption{Log of spectral power fractional bandwidth ($\delta\omega/\Omega_{ce}$) vs. electron beam kinetic energy. The colors indicate propagation direction, as indicated in right-hand color bar.}
  \label{fig-bw-kpar}
\end{figure*}

In order to compare with observed fractional bandwidths, we must consider the relation between growth rate and emergent spectral power. The spectral power is very sensitive to small changes in growth rate: For a typical e-folding amplification factor of 10 (section \ref{sec-rx-gain}), a 13\% reduction from peak growth rate corresponds to a 10 dB intensity decrease.  In addition, for non-perpendicular propagation, the resonant circle and the velocity distribution form osculating circles, so that very small changes in resonant circle radius result in large changes in growth rate.  Since the radius of the resonant circle is a function of fractional bandwidth  (equation \ref{eqn-radius}), the growth rate integrals are very sensitive to $\delta\omega$.  Figure \ref{fig-bw-kpar} shows the spectral power fractional bandwidth (defined as full width to -10 dB points) as a function of horseshoe energy for a range of propagation angles centered on $k_{\|} =0$. The majority of points cluster between $10^{-3.5}$ and $10^{-2.5}$, but at low energy and non-perpendicular propagation, some fractional bandwidths are as small as 10$^{-4.2}$. 
These fractional bandwidths encompass most observed narrowband AKR features, including the isolated striated AKR bursts reported by \citet{rlm06} who measured fractional bandwidths as small as $10^{-4}$. They are not quite as narrow as those reported by \citet{bc87} who reported fractional bandwidths as small as $3\times10^{-5}$. 

\subsection{RX-mode: Perpendicular propagation}
\begin{figure*}[p]
  \vspace*{2mm}
  \begin{center}
  \includegraphics[width=5.5in,angle=90]{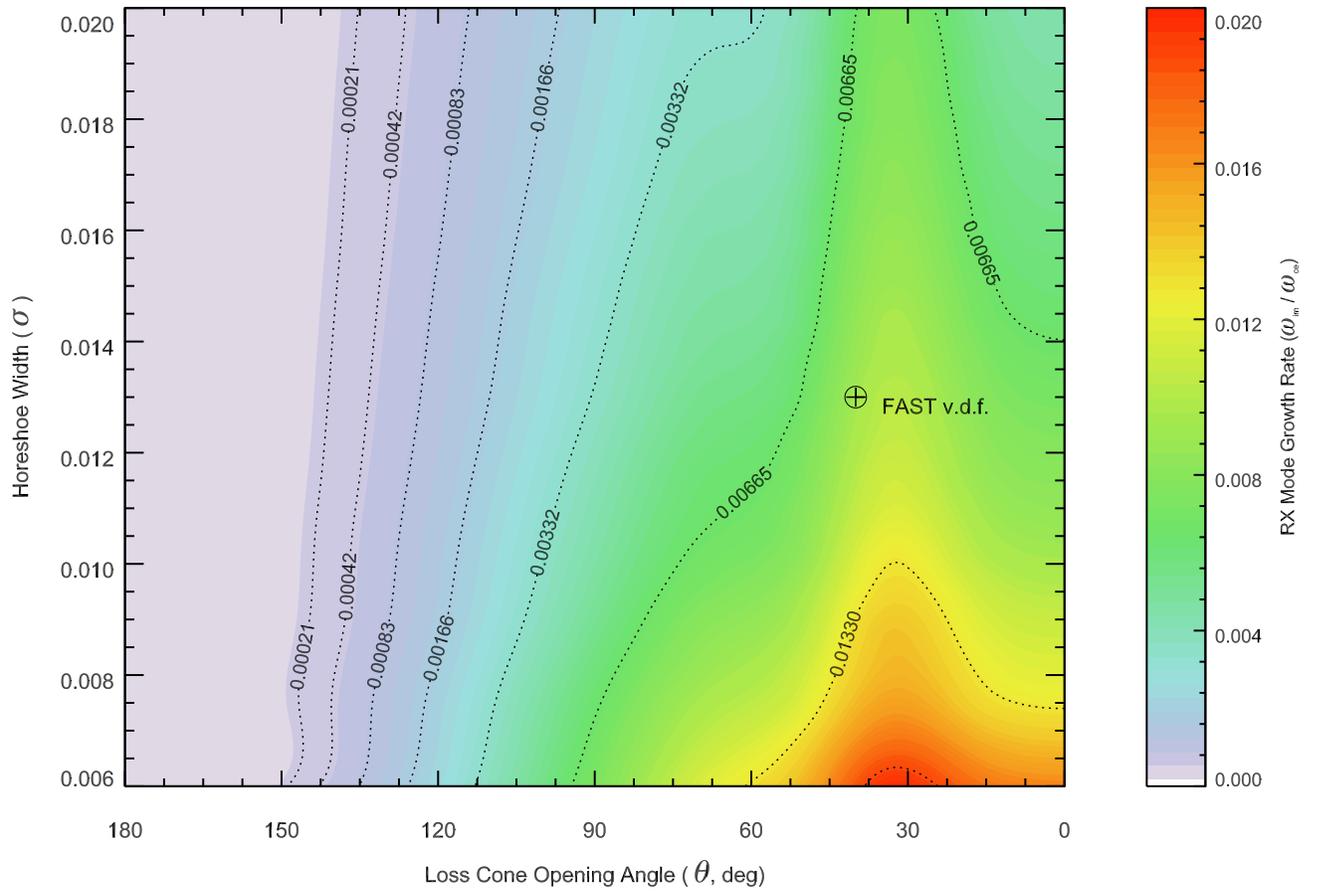}
   \end{center}
  \caption{RX mode growth rate vs. horseshoe width and loss cone opening angle at perpendicular propagation, with other parameters fixed at 'FAST match' values list in Table 1. Contours lines increase by a factor of two starting from 0.0015. The point labeled 'FAST v.d.f.' corresponds to the best-fit parameters matching the velocity distribution function described in \citet{m03}, FAST orbit 1804.}
  \label{fig-theta-sigma}
\end{figure*}

After inspecting the parameters associated with the highest growth rates, a number of trends became clear. First, we examine the dependence on loss-cone opening angle and horseshoe width. Figure \ref{fig-theta-sigma} shows the RX mode growth rate  as a function of horseshoe width and loss cone opening angle for the case $k_{\|}=0$ and with other model parameters fixed at the `FAST match' values in Table 1. For loss cone angles larger than $\sim90^{o}$  the growth rate decreases very rapidly,  indicating that the CMI mechanism requires broad pitch-angle scattering of the downward beamed electrons, resulting in a significant population of upward-traveling electrons and hence a horseshoe rather than a crescent geometry. 

The horseshoe width parameter, which is a measure of the energy spread of the beamed electrons, has a smaller effect on the growth rate. At the `FAST match' coordinates (labeled FAST v.d.f.), the width is 0.78 keV.  A width change of 0.01 (0.6 keV) corresponds to a growth rate change of  a factor of 2. 

\begin{figure*}[p]
  \vspace*{2mm}
  \begin{center}
  \includegraphics[width=5.5in, angle=90]{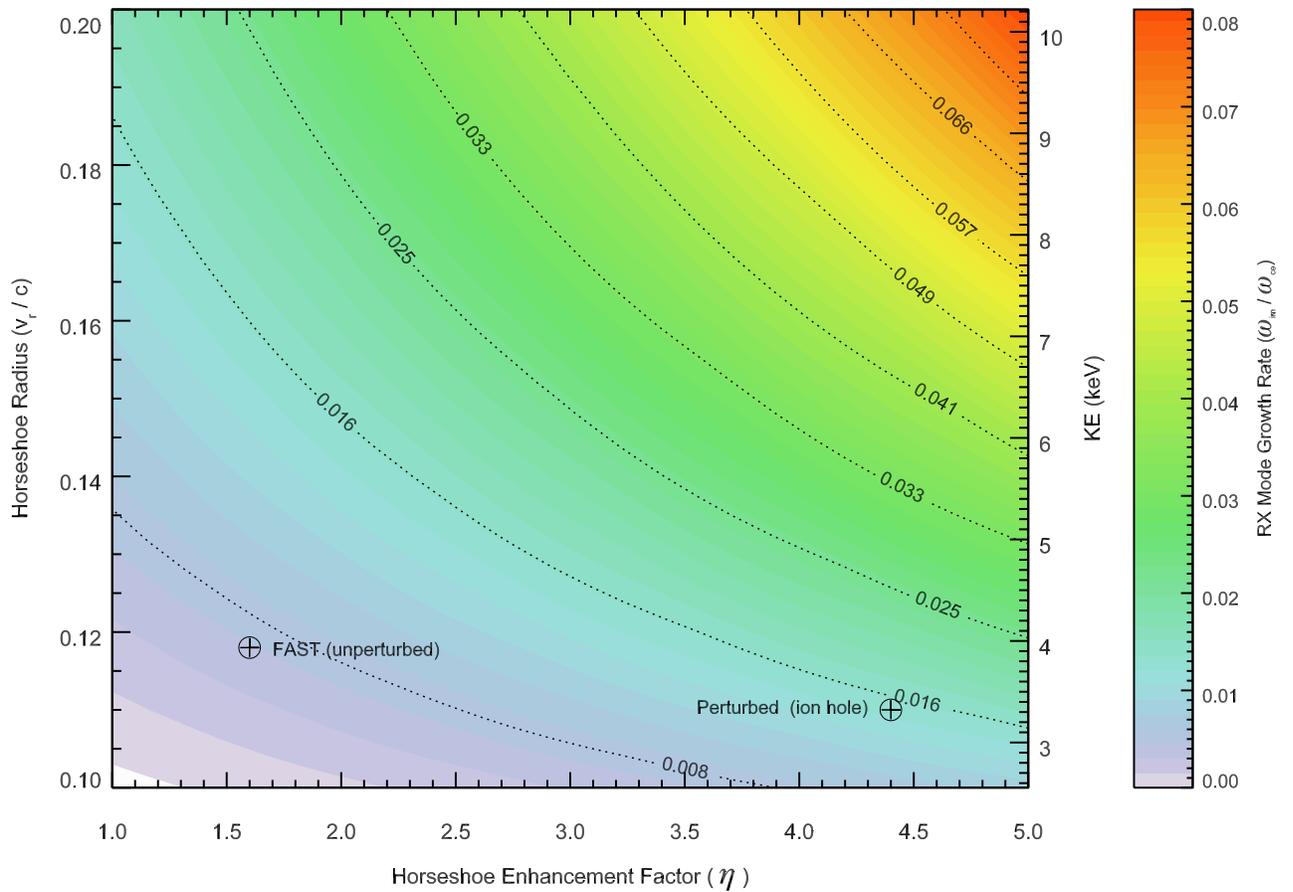}
 \end{center}
  \caption{Same as Fig. \ref{fig-theta-sigma} but for horseshoe radius $v_{r}/c$ and enhancement factor 
  $\eta$. Contour lines increase by 10\% starting at 0.0488. The point labeled 'FAST v.d.f. (unperturbed)' corresponds to the 'FAST match' parameters in Table 1. The point 'Perturbed (ion hole)'  corresponds to perturbation of the velocity distribution by a 700 eV ion solitary wave (Section 3}.
 \label{fig-vr-eta}
\end{figure*}

Figure \ref{fig-vr-eta} shows R-X mode growth rate as a function of horseshoe enhancement factor $\eta$ and radius $v_{r}$ with other parameters fixed at `FAST match' values as above. The right ordinate is the electron beam kinetic energy calculated from the horseshoe radius. As might be expected, the growth rate increases with increasing beam energy and $\eta$. The latter parameter characterizes the steepness of the horseshoe compared with the plateau component and is proportional to the partial derivative $\partial f /\partial v_{\bot}$ in the growth rate integral. 

\subsection{RX-mode: Non-perpendicular propagation}
\begin{figure*}[p]
  \vspace*{2mm}
  \begin{center}
  \includegraphics[width=5in, angle=90]{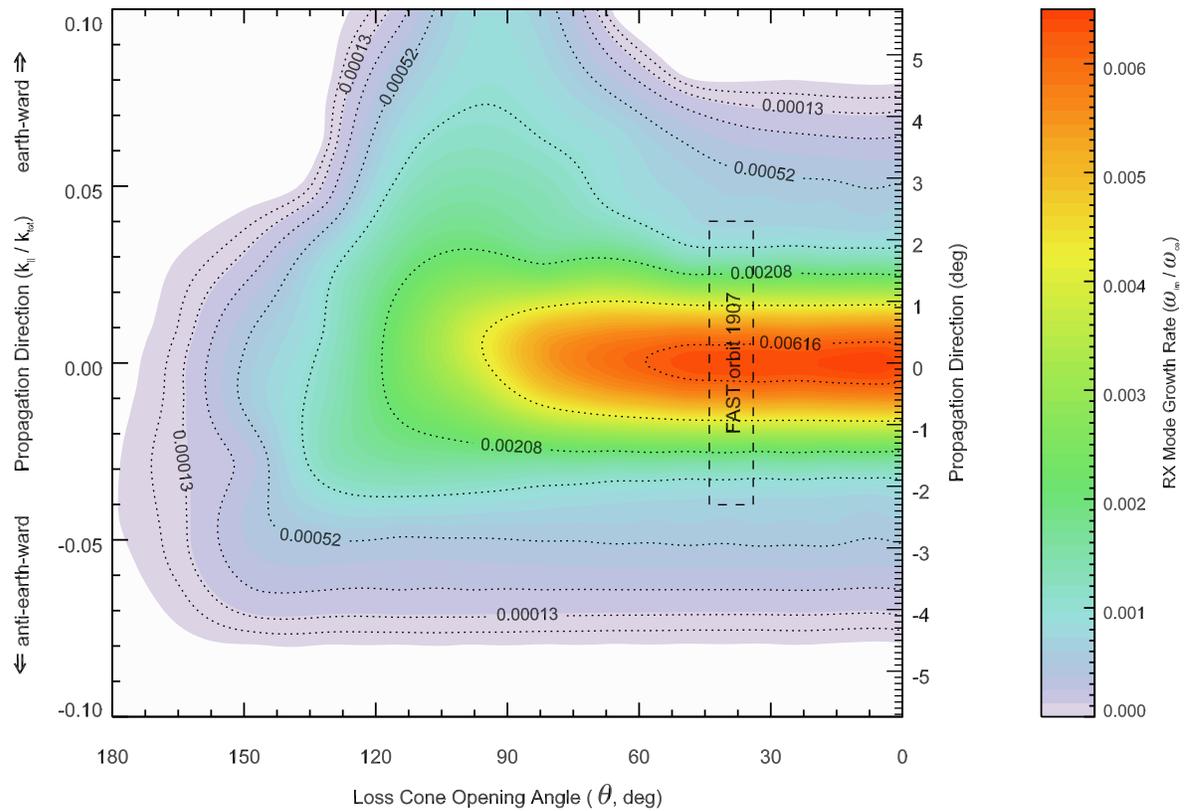}
 \end{center}
  \caption{ Same as Fig.\ref{fig-theta-sigma} but for loss cone opening angle $\theta$ and propagation direction $k_{\|}/k$. The dotted box labeled 'FAST orbit 1907' indicates the observed AKR propagation direction with uncertainties \citep{e98} and loss cone opening angle estimated from 2-dimensional distribution function \citep{d98}, FAST  orbit 1907, 13 Feb 1997 near 18:58:55 UT.
   }
  \label{fig-theta-kpar}
\end{figure*}

An illustrative case of  non-perpendicular propagation is shown in Figure \ref{fig-theta-kpar}. which  displays the RX-mode growth rate as a function of propagation direction  and loss cone opening angle, with other parameters fixed at `FAST match' values. The right ordinate displays propagation angle (measured from the normal direction) in degrees. The box labeled 'FAST orbit 1907' shows the estimated loss cone opening angle (with uncertainty) and k-vector propagation direction for FAST orbit 1907 observations on 13 Feb 1997 near 18:58:55 UT. The loss cone opening angle estimate was made using the distribution function shown in Fig. 2 of \citet{d98}, while the propagation direction and uncertainty was taken from \citet{e98} for the same time period. 

The growth rate is sharply peaked at $k_{\|}=0$ with a characteristic FWHM (full-width at half-maximum) $\pm 0.015$ , i.e., a propagation FWHM angle $1.7^{o}$. This agrees very well with measurements of k-vector direction observed by the FAST electric field sensors. There is also a small but significant bias toward the loss cone near  $\theta = 90^{o}$.  This is a result of the resonant circle intercepting  the inner shoulders of the upward loss cone.  While varying all other model parameters, we found no circumstances for which the propagation direction at peak growth deviated significantly from perpendicular. 

\subsection{LO-mode growth rate}

\begin{figure*}[p]
 \vspace*{2mm}
 \begin{center}
 \includegraphics[width=6.5in]{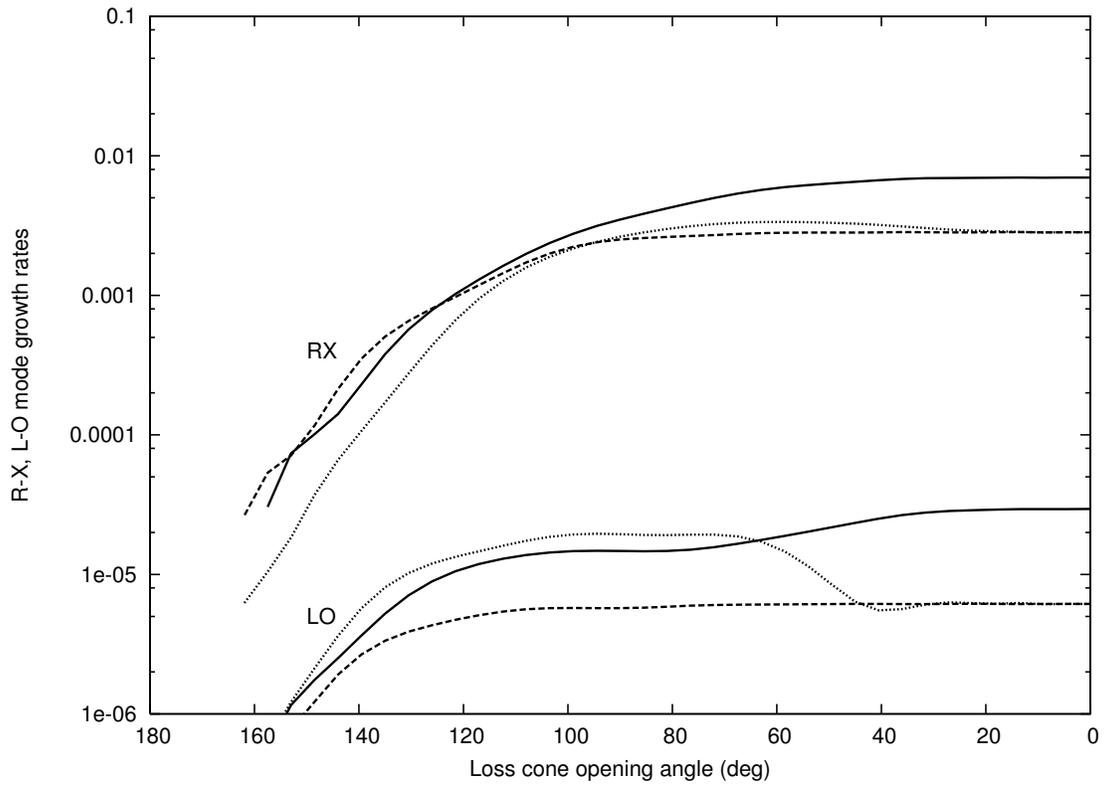}
 \end{center}
 \caption{Growth rate for R-X mode (filled circles) and L-O mode (open circles) versus loss cone opening angle with other model parameters fixed at 'FAST match' values. For each mode, the solid lines are for perpendicular propagation ($\phi =0^{o}$)  while the dashed and dotted lines are for $\phi = -1^{o}$ and $\phi =+1^{o}$ respectively. }
 \label{fig-rx-lo}
\end{figure*}

Figure \ref{fig-rx-lo}   shows growth rates for RX and LO mode emission as a function of horseshoe opening angle using fixed 'FAST match' for all other model parameters. The Figure shows the growth rates in each mode using three propagation directions: $k_{\|}/k\ =\  0^{o}$, (solid line), $1^{o}$  (dashed line), and $-1^{o}$ (dotted line). For these parameters, which are typical of the auroral cavity, the LO to RX mode growth rate ratio is $\sim10^{-3}$. We computed the ratio of RX to LO growth rates the over the entire range of parameters listed in Table 1. The ratio was less than $2\cdot10^{-3}$ for all  RX mode growth rates $\omega_i > 10^{-4}$.

We can understand this ratio qualitatively by inspection of the growth rate expressions (equations \ref{grow-rx}, \ref{grow-lo}). At a given density, the LO mode growth rate integrand differs only by the multiplicative term $(v_rsin\phi/c)^2=(v_{\|}/c)^2\sim\mathcal{O}(10^{-2})$ compared with the RX mode growth integrand. However, the remaining integrand heavily weights the line integral to regions near $v_{\|}\approx0$, resulting in a further reduction $\sim10^{-2}$. Finally, the RX-mode dispersion factor $\mu_R$ reduces the RX-mode growth rate relative to the LO-mode by $\sim10^{-1}$ (Appendix A). Hence, the expected overall ratio of RX to LO growth rates is $\sim10^{-3}$. 
\subsection{Comparison with observed AKR intensities}
\subsubsection{Group speed and convective growth length}
In order to compare calculated growth rates with observations of AKR intensities, we must calculate the expected emergent spectral power in each mode. To do this we need to know the physical length over which the CMI maser operates and the group speed of the wave. We must also determine the background intensity to determine the ratio of emergent flux to input flux at a given frequency.

The temporal growth rate $\omega_{i}$ and power gain $G$ (dB) of an unsaturated CMI maser are related by
\begin{equation}
G(dB) = 10\cdot 
\log_{10}
\left[
\exp{
\left(
2\omega_{i}\frac{L_c}{V_g}
\right)
}
\right]
\label{eqn-gain2}
\end{equation}

where $L_{c}$ is the convective growth length, i.e., the physical length over which the conditions for CMI growth are maintained, and $V_{g}$ is the group speed of the mode. A proper calculation the group speed of each mode requires solving the relativistic dispersion equation for  the assumed hot electron velocity distribution including a possible cold electron population which may also be present. This calculation is beyond the scope of the present paper. However, for the R-X mode, we can use the results of \citet{p84b} who calculated group speeds assuming a DGH distribution and a relativistic dispersion relation. He found that the group speed varied in the range  $0.2c>V_g>0.1c$  both for a range of DGH parameter $2<l<\infty$ and no cold electrons, and for the case $l=2$ and fractional cold electrons $0.0<n_c/n_h<0.75$.  We adopt an R-X mode group speed in the middle of this range, $V_g =0.15c$, although the uncertainty in this estimate is probably a factor of 2. For the LO mode, we assume $V_g =c$  since the radiated frequency is much larger than the LO cutoff frequency ($\omega_{pe}$). 

The convective growth scale is constrained by the physical dimensions of the auroral cavity. In the latitudinal direction the cavity size is $\sim$100 km or less \citep[e.g.,][]{b94}. In the direction along the tangent plane, the auroral cavity is much larger. However, the CMI mechanism requires very similar environmental  conditions along the convective growth path. Probably the most severe limitation is the change of magnetic field orientation with respect to the propagation vector. For a dipolar field a longitudinal shift of $1^{o}$ at an altitude of 4000 km corresponds to linear distance of 180 km. Inspection of Fig. \ref{fig-theta-kpar} shows that the growth rate is reduced by a factor of 2 for angular deviation of  $1^{o}$ from maximum growth. Hence we conclude that the convective growth scale is unlikely to exceed 200 km in the auroral cavity.

\subsubsection{RX Mode Gain \label{sec-rx-gain}}  

\citet{og82} estimated that the CMI maser required about 10 e-foldings ($e^{20}$ power gain, or 87 dB) to account for the 'observed AKR intensities' , a number that has been widely used in the AKR literature subsequently. However, more recent in situ observation of peak AKR emission electric field intensity in auroral cavities are now available which are significantly higher than previous estimates based on measured flux densities from remote spacecraft. Furthermore, since the sky background intensity and AKR peak intensities are highly frequency-dependent, we re-examine the required maximum gain and its possible frequency dependence. 

At frequencies between 200 KHz  and 800 KHz the sky background is dominated by synchrotron emission from the galactic plane. The emission peaks near 3 MHz and is strongly absorbed by free-free absorption from cold neutral hydrogen at lower frequencies \citep{b73}. The sky brightness is steeply rising with an approximate power-law form
\begin{equation}
B_{sky}(\nu) = B_{0} {\left( \frac{\nu}{\nu_{0}}  \right)}^{4}
\label{eqn-skybkgnd}
\end{equation}
where $B_{0} =5\ 10^{-22}$ W m$^{-2}$ Hz$^{-1}$ sr$^{-1}$ and $\nu_{0}=$ 360 Khz. This dependence is steeper than that expected for free-free absorption (exponent 5/2) and may indicate that Razin suppression by cold plasma in the synchrotron source region is significant.  Also, since the emission originates in the galactic plane, the background is not isotropic, but varies by a factor $\sim$2 over the sky at a given frequency. The constant $B_0$ represents an average over the maximum and minimum observed brightness at each frequency \citep{b73}. 

In order to compare with observed AKR spectral power, we convert the observed sky brightness to squared electric field spectral power \citep{b66,bf91}
\begin{equation} 
I_{sky}(\nu)= \frac{8\pi}{\epsilon_{0}c}B_{sky}(\nu)
\label{eqn-akr-e}
\end{equation}
where $I_{sky}$ is the averaged isotropic spectral power (V$^{2}$ m$^{-2}$ Hz$^{-2}$) and $\epsilon_{0}$ is the permittivity of free space. For AKR radiation measured at FAST orbital altitudes in the auroral cavity ($\nu\approx$ 360 KHz), the background flux is $5\ 10^{-18}$ (V/m)$^{2}$ Hz$^{-1}$. 

A number of papers have addressed the statistical properties of AKR spectral power as measured by remote spacecraft \citep[e.g.,][]{gg85,bf91,k98,g04}. However, in order to estimate the emergent spectral power using flux measurements from a remote spacecraft, the angular beaming pattern of AKR radiation must be known. This is currently not well established, so we use instead measurements of AKR spectral power from in situ spacecraft. The plasma wave tracker instrument on the FAST satellite, when operated in burst waveform mode \citep{e01}, allows direct measurement of AKR intensities within the auroral cavity. 

One of the most intense AKR bursts observed by FAST had a spectral power
$2\ 10^{-4}$ (V/m)$^{2}$  Hz$^{-1}$ \citep{e98}. We are not aware of a statistical analysis of AKR spectral powers observed by FAST, but \citet{e98} state that the average spectral power was `roughly one order of magnitude higher than that reported by Viking' \citep{f87}, which implies an average spectral power near $10^{-8}$ (V/m)$^{2}$ Hz$^{-1}$. Using these estimates and the background spectral power calculated above, the peak CMI gain at 360 KHz is 161 dB (18.5 e-foldings) and the average gain is 93 dB (10.7 e-foldings). The average is in good agreement with earlier estimates, although the peak gain is significantly larger.

\begin{figure*}[p]
 \vspace*{2mm}
 \begin{center}
 \includegraphics[width=5.5in,angle=90]{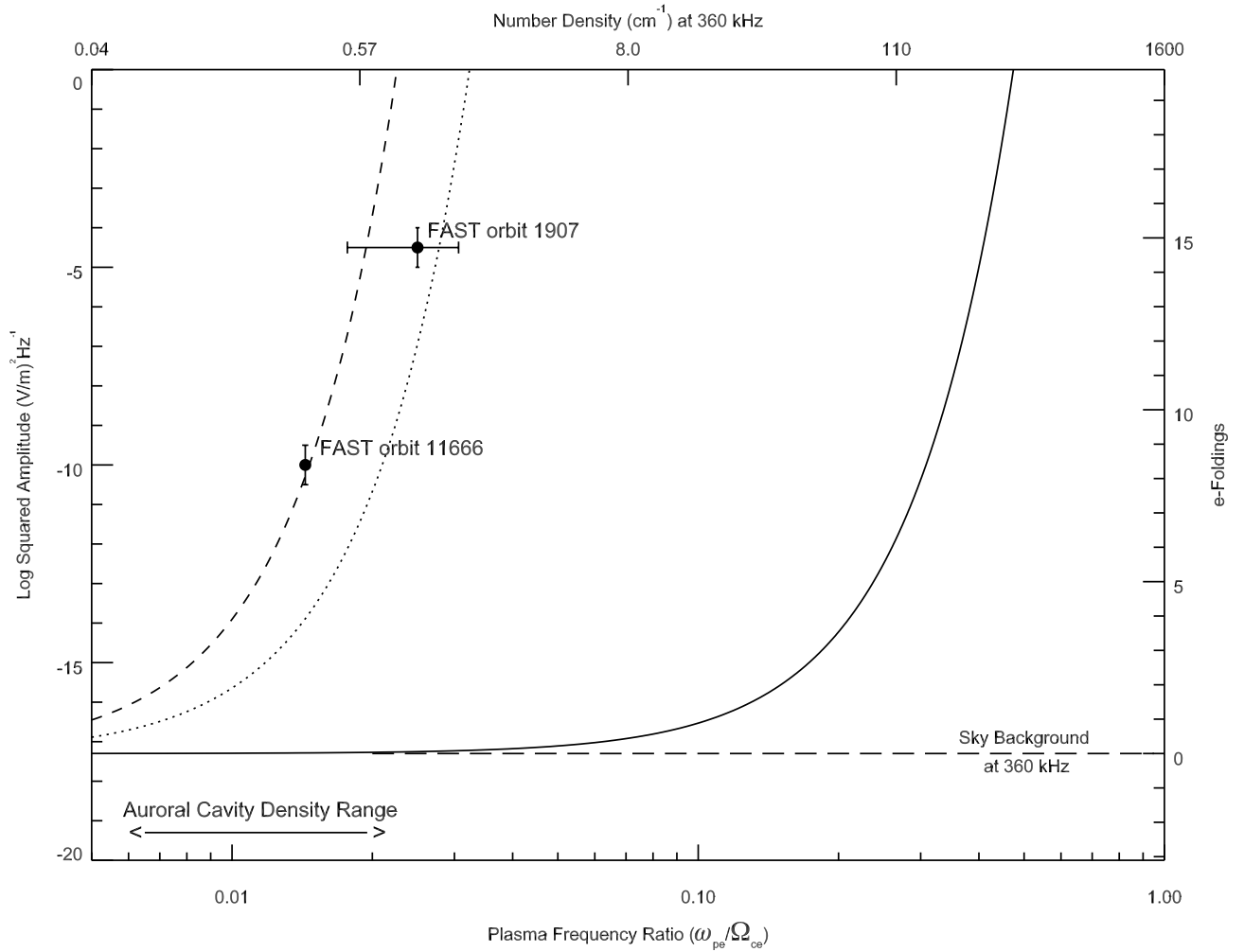}
 \end{center}
 \caption{Computed RX mode spectral power (left axis) and equivalent e-folding field amplification (right axis) of CMI maser as a function of plasma frequency ratio (bottom axis) and number density at  4000 km ($f_{ce} = 360$ KHz) using model parameters listed in Table 1 (FAST match column), $k_{\|} = 0$,  group velocity $V_g = 0.15c$ and convective growth length $L_c  = 35$ km (dashed line) and $L_c = 20$ km (dotted line). The points labeled FAST orbit 11666 and FAST 1907 are measured AKR spectral powers and electron densities (see text for details). 
The LO mode spectral power, using $V_g = c$ and $L_c= $200 km (solid line), does not approach observed intensities until the electron number density is more than $100\times$ the observed mean auroral cavity density. }
 \label{fig-gain-efold}
\end{figure*}

Figure \ref{fig-gain-efold} is a plot of the calculated RX-mode spectral power versus plasma frequency ratio for $k_{\|}=0$, group speed $V_g = 0.15c$, and convective growth lengths $L_c =  20$ km (dotted line) and 35 km (dashed line). We varied the density parameter $F_0$, but fixed the remaining parameters using 'FAST match' values (Table 1).  The point labeled 'FAST orbit 1907' indicates the observed spectral power and electron density (with uncertainty) for the intense burst observed by FAST on 13 Feb 1997 at 18:58:55 UT \citep{e98}. The point marked 'FAST orbit 11666' is an AKR burst observed by FAST on 2 Aug 1999 at 23:33:52, and using the density measurement reported in \citet{m03} and spectral power  of the AKR burst measured from the FAST data archive. The observed AKR spectral powers agree very well with the model, although the convective growth scales should be considered minimum estimates, since any deviation from normal propagation reduces the growth rate considerably, necessitating larger convective growth lengths.

\subsubsection{LO mode gain and RX/LO intensity ratios}

We next investigate the question of whether a parameter search of calculated LO mode growth rates can yield ratios  of RX to LO growth rates which are consistent with observed values. When LO mode is detected, it is often a few percent of the RX mode intensity \citep{m84, b88} although it occasionally reaches 100\%  \citep{h03}.  

Figure \ref{fig-gain-efold}  shows the LO mode gain using 'FAST match' parameters an assumed for group velocity $V_g = c$ and $L_c = $ 200 km (solid line) ), which we have argued (section 2.8,1) is the largest plausible convective growth length near 4000 km altitude.  The gain does not approach observed levels (8-10 e-foldings) unless the plasma ratio $\omega_{pe}/\Omega_{ce}\geq0.3$.  This implies energetic electron number densities at least 100 times larger than observed in the aurora cavity.
This suggests that LO mode AKR radiation, at least inside auroral cavities, is not generated directly by the CMI mechanism. Simulations by \citet{ps85} suggest that a relativistic effect coupling X-mode (perpendicular) electric fields into parallel (O-mode) fields may account for the observed LO to RX-mode intensity ratios. The LO mode may also be created from conversion of RX mode emission by polarization mismatch in regions of high density gradients \citep{hm86,l96b}. 

\section{Growth rate modification caused by solitary structures and EMIC waves}

We next consider change of CMI growth rate and consequent gain change caused by modification of an initial horseshoe distribution function within solitary structures (a.k.a. waves or holes). Solitary structures have recently been suggested as the cause of frequency fine structure seen in dynamical spectra of AKR radiation \citep{ptb01,ptbj03,pt05,rlm06}. We also briefly consider the effect of electromagnetic ion-cyclotron (EMIC) waves, which have been associated with stimulation of AKR \citep{men06,x07}.

\begin{figure*}[p]
  \vspace*{2mm}
  \begin{center}
 \includegraphics[width=8in,angle =90]{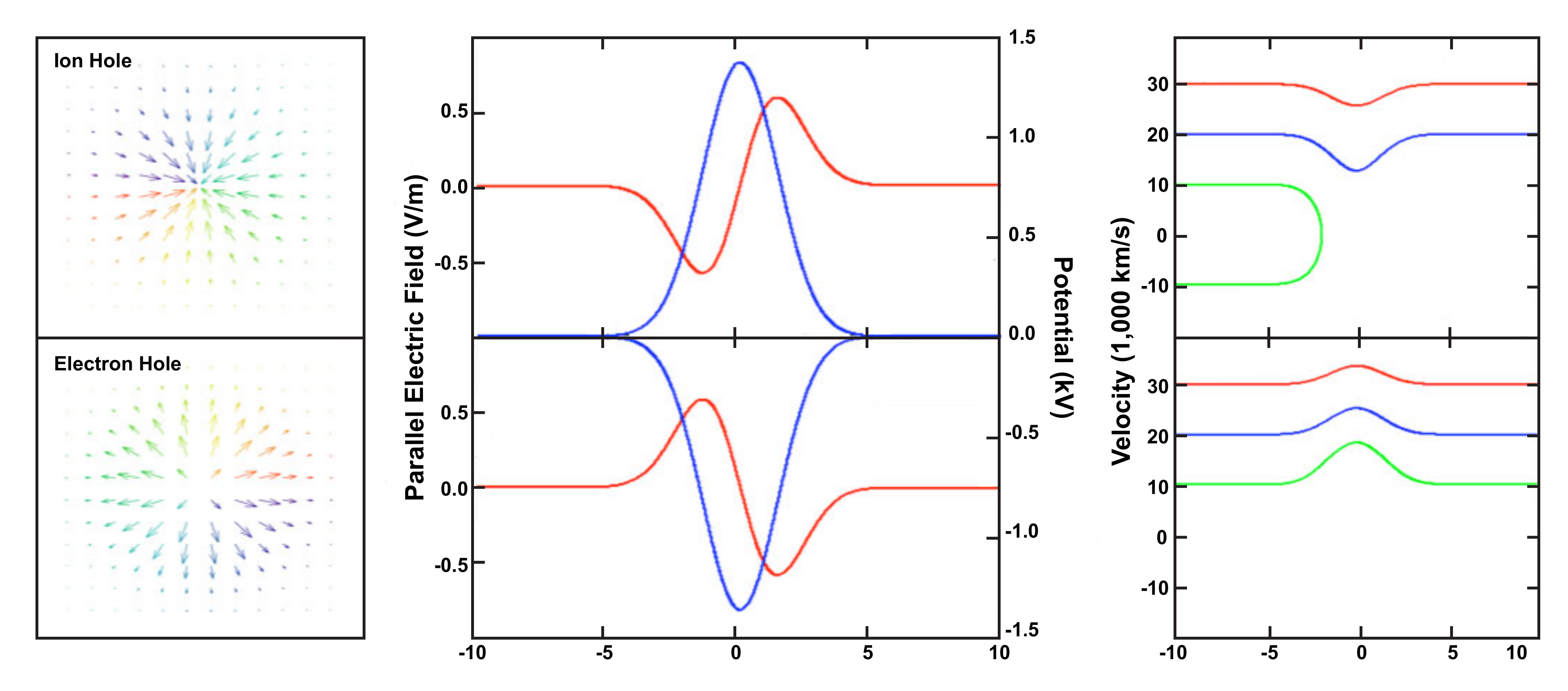}
   \end{center}
  \caption{Effect of ion and electron holes on velocity distribution function. ; $(a)$ Model E-field vectors for spherical ion hole (top) and electron hole (bottom); $(b)$ Velocity profile versus distance from hole center for incident electrons with initial with speeds of 30,000 km s$^{-1}$ (red), 20,000 km s$^{-1}$ (blue), and 10,000 km s$^{-1}$ (green); $(c)$ perturbed velocity distributions modified by ion hole and electron holes with a total potential of 0.7 keV.}
 \label{fig-iss-cartoon}
\end{figure*}

We assume a simple analytic model for electron and ion solitary structures as spherically symmetric potential structures with an  electric field structure as shown in Figure \ref{fig-iss-cartoon}. The electron hole accelerates the parallel component of all electrons passing the structure, while the ion hole causes a reduction in the parallel speed (in both cases, the perpendicular component is largely unaffected since the gyroradius is much smaller than the size of the hole). The parallel velocity perturbation at the center of the hole can estimated by differentiating the energy conservation equation

\begin{equation}
\delta V_{\|}= \pm\frac{q_{e}\delta\Phi}{m_{e}V_{\|}}
\label{eqn-deltav}
\end{equation}

where $\delta\Phi$ is the effective hole potential, and the perturbation is negative for ion holes and positive for electron holes. This local deformation of the distribution function can either enhance or diminish the CMI growth rate depending on whether the deformation shifts the perpendicular velocity gradient closer to or away from the resonant circle. Even if the effect is to enhance the growth rate, the spatial scale of the deformation must be sufficiently large so that the CMI gain is significant over the structure.
In order to investigate the full range of possible conditions, we calculated the RX mode growth rates inside and outside the structures, varying all model distribution parameters as well as propagation direction. 

\subsection{Perturbation by Ion holes}

Ion solitary waves (ion holes) are observed in the upward current region as symmetric bipolar  electric field structures with negative potential. They are often seen moving upward at several hundred km s$^{-1}$ in association with upgoing ion beams \citep{m03}. They have spatial scale sizes 2-12 km and typical amplitudes $E\sim10$ mV/m, although large structures can be several hundred mV per meter. Interior to such large structures, the parallel component of the velocity distribution function will be shifted significantly toward the origin, by up to 1 keV in energy,  corresponding to a 5\% - 15\% decrease in parallel velocity for typical range of energetic electron energies. 
 
 We model this perturbation  by adding a velocity-dependent term to the distribution function
 \begin{equation}
 \delta V_{\|} = -\kappa \frac{c^2}{V_{\|}} 
 \label{eqn-kappa}
 \end{equation}
 where $\kappa$ is a (dimensionless) perturbation parameter.  Electrons with kinetic energies less than the hole potential will suffer parallel velocity reversal rather than the decrement given in equation \ref{eqn-deltav}, as illustrated in Fig. \ref{fig-iss-cartoon}b. 
 
\begin{figure*}[p]
  \vspace*{2mm}
  \begin{center}
 \includegraphics[width=6.5in,angle =0]{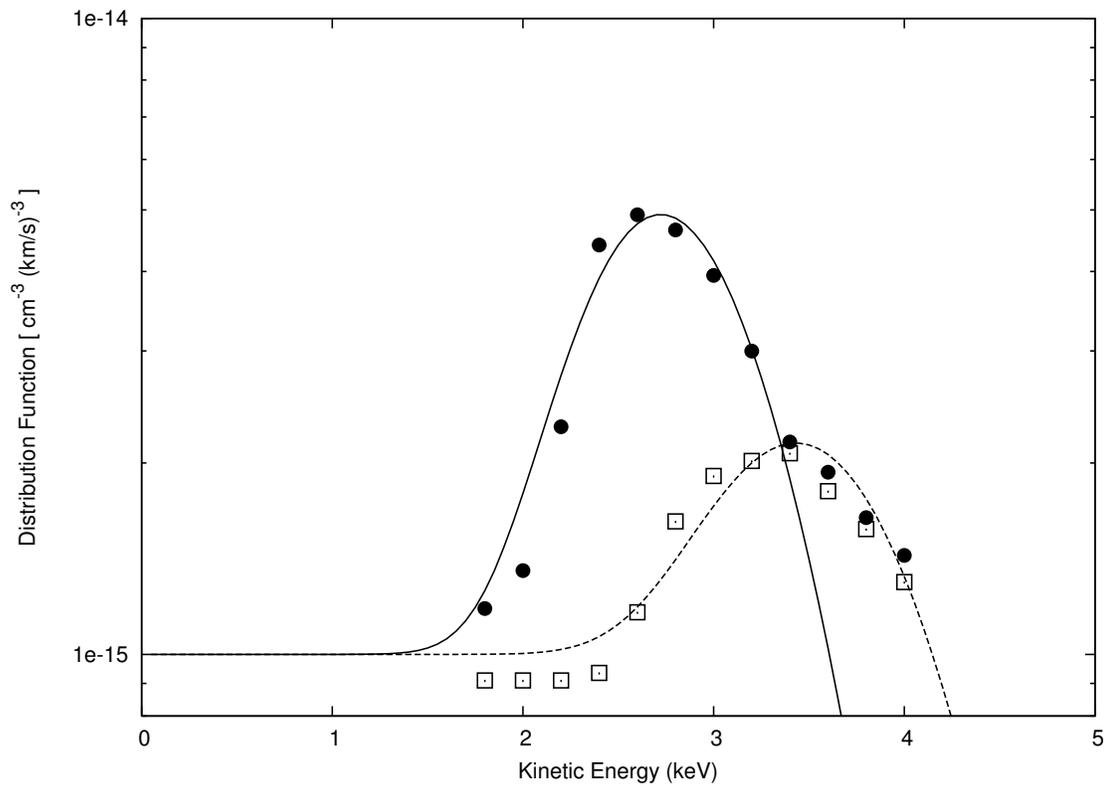}
   \end{center}
  \caption{One-dimensional radial profile of model distribution function using 'FAST match' parameters (dashed line). Open squares are FAST observations scaled from one-dimensional distribution function plotted in \citet{m03} Fig. 11 (lowest trace).  The perturbed distribution profile (solid line) corresponds to  velocity best-fit perturbation parameter $\kappa = 0.014$ and $\eta$ = 4.4, The solid circles are data points scaled from the middle trace of \citeauthor{m03} when the perturbation from a passing ion solitary structure is maximal.}
  \label{fig-model-iss}
\end{figure*}

 A detailed, self-consistent calculation of the deformed distribution function involves solving the Vlasov equation given the potential structure of the hole, which is beyond the scope of this paper.  Instead, we adopt a heuristic approach by adjusting the horseshoe density shoulder factor $\eta$ to fit the observed distribution function within an ion hole. This approach is justified by the close agreement between the model distribution and the observed distribution within an ion solitary wave.
 An excellent example of a distribution function perturbed by the passage of a large (700 eV) ion solitary wave in shown in Fig. 11 of \citet{m03}.  The figure shows a time-sequence of  one-dimensional distribution function profiles (averaged over $112^{o} - 180^{o}$ pitch angle) as the solitary wave passed by the FAST spacecraft. The sequence clearly shows a 700 eV energy shift toward the origin, comparable to the total potential of the wave, as well as a $3\times$ increase in the peak amplitude. Figure \ref{fig-model-iss} show sampled points from this time-series for the unperturbed distribution (open squares) and for the maximally shifted distribution (filled circles). The dashed line is the best-fit model using `FAST match' parameters (cf. Fig. \ref{fig-2x1-vdf}b), while the solid line was  obtained by adjusting $\kappa$ and $\eta$ and retaining 'FAST match' values for all other parameters. We used this perturbed model distribution to calculate the growth rate inside and outside the solitary wave (see labeled points,  Fig. \ref{fig-vr-eta})
  
\begin{figure*}[p]
 \vspace*{2mm}
 \begin{center}
 \includegraphics[width=6.5in,angle=0]{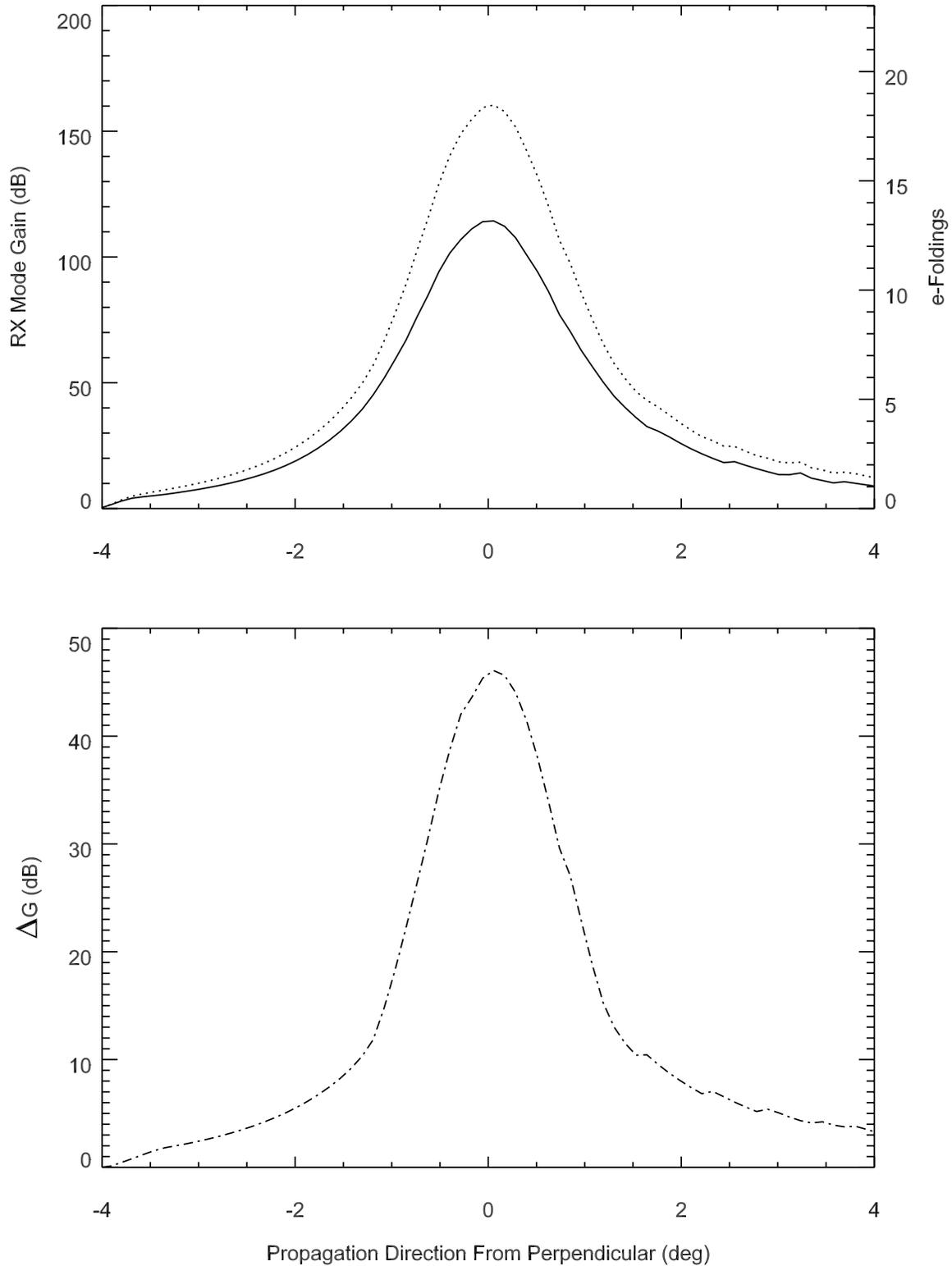}
 \end{center}
 \caption{$(a)$ R-X mode gain (left axis) and corresponding e-foldings (right axis) versus propagation direction for horseshoe velocity distribution function with 'FAST match' parameters list in Table 1 assuming a 40 km convective growth path and group velocity $V_g = 0.15c$. The solid line is for the unperturbed distribution, while the dashed line is for a 0.7 keV ion hole and the dotted line is a 0.7 keV electron hole. $(b)$ As above, but gain difference between unperturbed distribution and perturbation caused by ion hole. }
 \label{fig-gain-iss}
\end{figure*}

 We now assume the effective convective growth length is $1/e$ the overall spatial scale of the solitary wave ($L_c\sim$4 km for large structures), and a group speed $V_g\approx0.15c$, as discussed in section 2.8.  Figure \ref{fig-gain-iss}(a) shows the resulting RX mode power gain as a function of propagation direction for the unperturbed state (solid line), and for the perturbed distribution within the ion hole (dashed line). The lower plot (Fig. \ref{fig-gain-iss} $(b)$) shows the gain difference, which exceeds 40 dB over the central part of the gain curve. This is comparable with the intensity contrast  of the brightest striated AKR bursts compared with the background AKR as reported by \citet{rlm06}, which were ascribed to perturbation by ion holes.
 
 \subsection{Perturbation by Electron holes}
 Electron holes (also called electron solitary structures or fast solitary waves) are isolated regions of positive potential with large electric fields, up to 2.5 V/m. They are normally found in the downward current region \citep{e98b}. However, \citet{pt05} argue that the presence of 'tripolar' structures in the upward current region provides evidence for nested ion and electron holes in this region, similar to numerical simulations \citep{g03}, and that the electron holes are the cause of AKR fine structure. 
 
In order to determine whether electron holes can modify AKR growth rates, we calculated the change in growth rate caused by an single electron hole using the same technique as the ion hole calculation, viz.,, a spherical distribution (Fig. \ref{fig-iss-cartoon}b) in which incoming electrons are accelerated with a a velocity-dependent change given by equation \ref{eqn-deltav}, but with a positive sign. We determined the growth rate change for the entire range of all parameters listed in Table 1. Most parameters produced either a negligible difference or a net reduction in growth rate, up to 30\%. Even in cases of substantial reduction in growth rate, the change in emergent spectral power was negligible since the overall size of electron holes  ($\sim$0.5 km, \cite{e98b}), and hence the convective growth scale, is too small to substantially affect emergent power.  We conclude that electron holes are very unlikely to significantly modify AKR emission intensities.
 
 \subsection{Perturbation by EMIC Waves}

Electromagnetic ion-cyclotron (EMIC) waves have many properties similar to ion holes: They are frequently detected in the auroral cavity, they have peak amplitudes up to $\sim0.5$ keV, and spatial scales (wavelength) 1-10 km \citep{c98}. Similar to ion holes, they also have been observed to modulate the ambient electron energy distribution with an amplitude comparable to the total EMIC potential  \citep{m98}. \citet{men06} and \citet{x07} suggest that EMIC waves could stimulate AKR emission by modulation of the electron phase space density. 

From this perspective, an EMIC wave train can be considered as a string of ion holes separated by positive potential structures of equal amplitude and spatial dimension. Since we have shown that ion holes can increase the AKR spectral power by up to 40 dB, while positive potentials decrease the gain, the net result for each wave crest would be the same as an ion hole of similar amplitude, but with a 50\% duty cycle, i.e., a 3 dB decrease relative to a similar ion hole. The aggregate effect will would result in the same gain enhancement but spread over a much larger frequency range. Although there are large variations in EMIC wave packet sizes and direction, if we  assume a wave train has of order 10$^2$ crests, a wavelength of order 1- 10 km, and a k-vector direction close to the magnetic field direction, the resulting z-direction spatial width is $10^2 - 10^3$ km. At a mean height of 4000 km this results in an effective bandwidth 0.4 - 4 KHz. This is similar to the bandwidths of many drifting AKR bursts seen in AKR dynamical spectra \citep[e.g.,][]{g79,mm02}

\section{Summary}

We have used a parameterized horseshoe electron velocity distribution function to study the 
dependence of RX and LO-mode cyclotron maser instability  growth rates on horseshoe geometry, electron density, and propagation direction. The distribution function has been closely matched to distributions observed by the FAST spacecraft in the auroral cavity . We find that growth rates are most strongly affected by loss-cone opening angle and propagation angle. For loss cone opening angles exceeding $90^{o}$ the growth rates drop dramatically, indicating that significant AKR emission only occurs in regions for which pitch angle scattering of the downward electron beam produces upward traveling electrons. The dependence on propagation angle is sharply peaked at $k_{\|} =0$ with a characteristic full  width at half maximum of $1.7^{o}$. The fractional bandwidth of CMI emission, which  has a complicated dependence on horseshoe geometry and propagation angle, ranged from $10^{-4.3}$ to $10^{-2}$, but with a concentration near $10^{-3}$.  The range incorporates almost all observed AKR burst bandwidths, including the very narrowband striated bursts reported by \citet{rlm06}. 

We calculated the expected emergent spectral power for each mode given the growth rates, frequency-dependent sky background, and estimates for the convective growth scales and group velocities. The resulting spectral powers for RX-mode matched observed FAST observations of AKR power with convective growth scales of order 10 km and propagation angles $1-2^{o}$. The LO-mode spectral power is far lower than observed values even with a convective growth scale as large as 200 km, which we estimate to be the largest viable convective growth length in the auroral cavity. By performing a systematic parameter search, we find that LO-mode spectral power is consistent with observed values only for $\omega_{pe}/\Omega_{ce} > 0.2$, i.e., energetic number densities $n_e> 50$ cm$^{-3}$ at 4,000 km altitude, far greater than observed values. We conclude that LO-mode in the Earth's magnetosphere is probably not generated directly, but arises from mode conversion of RX-mode radiation.

Electron velocity distributions are perturbed by passing solitary structures (ion and electron holes) and by electrostatic and electromagnetic waves. We investigated how the RX-mode growth rate is affected by an ion hole, by model-fitting FAST measurements of the perturbed velocity distribution caused by an ion hole. We find that large (700 V) ion holes can enhance the CMI spectral power by up to 40 dB, with the maximum gain near $k_{\|} =0$. Similarly , we suggest that this result can be applied to intense electromagnetic ion-cyclotron (EMIC) waves, which act as a train of ion holes. The resulting bandwidth of EMIC-enhanced emission is much larger than for an ion hole, typically a few kHz. We used a similar analysis for electron holes, searching all ranges of distribution function parameters. Since all growth rates with electron hole perturbations were either unchanged or smaller than those without a perturbation, we conclude that electron holes cannot be 'elementary radiation structures'  \citep{ptb01, ptbj03}.

\section{Appendix A: Correction for relativistic dispersion}
The growth rate integrals (equations 2, 3, with $\mu_{R,L}$=1) were first derived by \citet{w79}, assuming that plasma dispersion could be neglected (refractive index $n=1$). For the ordinary mode and normal propagation, this is a good approximation, since the dispersion relation for mildly relativistic, zero-temperature plasma only differs from the cold-plasma  ordinary mode by the relativistic mass \citep{p84b}
\begin{equation}
\omega^2 = (kc)^2 + \frac{{\omega_{pe}}^2}{\gamma}
\end{equation}
where $\gamma$ is the Lorentz factor of the energetic particles. For  $\gamma\sim1$ and $\omega_{pe} << \Omega_{ce}\sim\omega$ the refractive index $n_L \sim1$, so we can set $\mu_L = 1$.

The RX-mode dispersion case is more complicated. \citet{lw80} generalized the derivation of \citeauthor{w79} to include dispersion by a dominant population of cold electrons. \citet{owg84} showed that the  RX-mode growth rate equation could be recast, including dispersion, by adding a dimensionless multiplicative term in front of the resonant integrals
\begin{equation}
\mu_R =\frac{2\eta}{\left[\omega\frac{\partial}{\partial\omega}{\rm Re(\Lambda)}\right]_{\omega=\omega_r}}
\label{eqn-mu}
\end{equation}

where the function $\eta(\omega,\vec{k},\Omega_{ce})$, and the dispersion matrix $\Lambda$ are given in their Appendix A. They evaluated this term assuming that the hot electrons were a small fraction of the cold background plasma. They found in a large reduction in the growth rates, up to to two orders of magnitude relative to growth rates in the absence of dispersive corrections. They concluded that their derived growth rates were too low explain observed AKR intensities.

More recently,  spacecraft particle measurements in the auroral cavity have shown that the cold plasma assumption is invalid:  The hot, mildly relativistic component dominates, and relatively few (if any) cold electrons are present  \citep{d98}.  In addition, in the cold-plasma formalism, RX-mode perpendicular waves cannot propagate (since the cutoff frequency lies above the gyrofrequency), whereas in the auroral cavity the most intense radiation is perpendicular. 

In order to solve for the RX-mode growth rate self-consistently, one must either solve equation (\ref{eqn-mu}) using a specified distribution function and the relativistic form of the dispersion matrix  or appeal to numerical simulations which incorporate the correct relativistic dispersion relations. \citet{p86} used such a numerical model, and found that  the RX-mode growth rates are much larger than those computed using a cold-plasma dispersion assumption, although still significantly smaller than growth rates computed in absence of dispersion. 

In this paper we are primarily interested in the dependence of growth rate on parameterized velocity distributions.  The conductivity tensor, and hence the dispersion matrix, is a function of integrals over the distribution function, while the growth rate is most sensitive to the topology of the distribution function itself, especially the locations of positive velocity gradients with respect to the resonance ellipse.  Hence to first order we can consider the dispersion correction as a scale factor whose value depends only weakly on 
the model parameters. We 
have estimated this scale factor by comparing the growth rate computed using the resonant integral formalism with the numerical simulations obtained by \citet{p84b} using the same distribution function, viz.,  a DGH distribution with parameter range $2<l<10$ and perpendicular propagation.  The growth rate ratio for the two calculations gave a dispersion correction factor $\mu_R= 0.14\pm0.03$ where the uncertainty corresponds to the ratio variation with parameter $l$.  

In principle this factor could be different with horseshoe distribution functions. However, the excellent agreement of RX-mode growth rates computed in this paper using the derived scale factor  $\mu_R$ with numerical simulations using  similar horseshoe distributions \citep{ps85, p86, p99} indicate that the dispersion factor does not depend strongly on details of the velocity distribution geometry.

\begin{acknowledgements}
The authors thank P. Pritchett for several very useful discussions. This research is supported by NASA grant XXX to the University of Iowa.
\end{acknowledgements}


\end{article}

\clearpage
\begin{table}[t]
\caption{Model Parameters  and Search Ranges}
  \begin{center}
    \begin{tabular}{llcccl}
  \tableline
   Model Parameter & Symbol & FAST match\tablenotemark{a}  &FAST value\tablenotemark{b} &Range & Range Notes\\
  \tableline
    Density normalization & $F_{0}$ & 10$^{-18}$ & 0.32 cm$^{-3}$ &$10^{-19} - 10^{-17}$ & Units: s$^{-3}$m$^{-6}$ \\
   Horseshoe radius & $v_r$ &  $0.118c $& 3.5 keV &$ 0.10c - 0.20c$ & $2.5 < KE < 10.2$ keV \\
   Horseshoe width  & $\sigma $ & $0.013c$ & 0.78 keV & $ 0.01c - 0.02c$ & \\
   Loss-cone opening angle & $\theta$ & 40$^{o}$ & - &$0^{o} - 180^{o}$ & 
   Full width is $2\times\theta$ \\
  Horseshoe shoulder & $\eta$ & 1.6 & -& $1.0 - 3.0$ & \\
  Parallel wavenumber ratio & $k_{\|}/k$ & - & 0\tablenotemark{c} &  $-0.1 - +0.1$ & \\
  \tableline
  Perturbation parameter &  & & & & \\
  \tableline
  Energy perturbation factor & $\kappa$ & 0.014 & - & -& Eqn. \ref{eqn-kappa}\\
  Shoulder perturbation factor & $\eta\prime$ & 4.4 & - & -& see FAST match Fig. \ref{fig-model-iss}\\
  \tableline
   \end{tabular}  \end{center}
\tablenotetext{a}{Parameters which best match observed FAST distribution function shown in \citet{m03}}
\tablenotetext{b}{Physical value corresponding to FAST model parameter.}
\tablenotetext{c}{Measured AKR k-vector direction , 0.06  but consistent with $k_{\|}=0$,  \citet{e98}, FAST orbit 1907.}
\tablenotetext{d}{Measured ion solitary wave amplitude \citet{m03}, FAST orbit 11666.}
\end{table}

\end{document}